\documentclass[aps,prl,twocolumn,superscriptaddress,10pt,floatfix,amsmath,amssymb]{revtex4-2}

\usepackage{graphicx} % Required for inserting images
\usepackage{dcolumn}% Align table columns on decimal point
\usepackage{xcolor}
\usepackage{bm}
\usepackage{physics}
\usepackage{lipsum}  
\usepackage{hyperref}
\usepackage[normalem]{ulem}

\graphicspath{{images/}}

\newcommand{\sect}[1]{\textit{#1}.-- }

\newcommand{\bth}{\bm{\theta}}

\begin{document}

\title{Shallow quantum circuits are robust hunters for quantum many-body scars}

\date{\today}
    
\author{Gabriele Cenedese}
    \email{gcenedese@uninsubria.it}
    \affiliation{Center for Nonlinear and Complex Systems, Dipartimento di Scienza e Alta Tecnologia, Universit\`a degli Studi dell'Insubria, via Valleggio 11, 22100 Como, Italy} 
    \affiliation{Istituto Nazionale di Fisica Nucleare, Sezione di Milano, via Celoria 16, 20133 Milano, Italy}

\author{Maria Bondani}
    \email{maria.bondani@uninsubria.it}
    \affiliation{Istituto di Fotonica e Nanotecnologie, Consiglio Nazionale delle Ricerche, via Valleggio 11, 22100 Como, Italy}
    
\author{Alexei Andreanov}
    \email{aalexei@ibs.re.kr}
    \affiliation{Center for Theoretical Physics of Complex Systems, Institute for Basic Science (IBS), Daejeon - 34126, Korea}
    \affiliation{Basic Science Program, Korea University of Science and Technology (UST), Daejeon 34113, Republic of Korea}

\author{Matteo Carrega}
    \email{matteo.carrega@spin.cnr.it}
    \affiliation{CNR-SPIN,  Via  Dodecaneso  33,  16146  Genova, Italy}

\author{Giuliano Benenti}
    \email{giuliano.benenti@uninsubria.it}
    \affiliation{Center for Nonlinear and Complex Systems, Dipartimento di Scienza e Alta Tecnologia, Universit\`a degli Studi dell'Insubria, via Valleggio 11, 22100 Como, Italy} 
    \affiliation{Istituto Nazionale di Fisica Nucleare, Sezione di Milano, via Celoria 16, 20133 Milano, Italy}
    \affiliation{NEST, Istituto Nanoscienze-CNR, I-56126 Pisa, Italy}
    
\author{Dario Rosa}
    \email{dario\_rosa@ictp-saifr.org}
    \affiliation{Center for Theoretical Physics of Complex Systems, Institute for Basic Science (IBS), Daejeon - 34126, Korea}
    \affiliation{ICTP South American Institute for Fundamental Research \\
    Instituto de F\'{i}sica Te\'{o}rica, UNESP - Univ. Estadual Paulista \\
    Rua Dr. Bento Teobaldo Ferraz 271, 01140-070, S\~{a}o Paulo, SP, Brazil}

%%%%%%%%%%%%%%%%%%%%%%%%%%%%%%%%%%%%%%%%%%%%%%%%%%%%%%%%%%%%%%%
\begin{abstract}

Presently, noisy intermediate-scale quantum computers encounter significant technological challenges that make it impossible to generate large amounts of entanglement.
We leverage this technological constraint as a resource and demonstrate that a shallow variational eigensolver can be trained to successfully target quantum many-body scar states.
Scars are area-law high-energy eigenstates of quantum many-body Hamiltonians, which are sporadic and immersed in a sea of volume-law eigenstates.
We show that the algorithm is robust and can be used as a versatile diagnostic tool to uncover quantum many-body scars in arbitrary physical systems.
\end{abstract}

\maketitle

\sect{Introduction}
The expectations and promises of the \textit{second quantum revolution}~\cite{Dowling2003,qcbook} are leading to huge resources invested in studying how to build new technological tools based on quantum mechanical principles, collectively dubbed as \textit{quantum technologies}. 

Among all, a dominant role is played by quantum computing and the recent successes and results obtained via the so-called noisy intermediate-scale quantum (NISQ) computers and the associated \textit{hybrid algorithms}~\cite{mcclean2016,cerezo2021, bharti2022,lau2022}.
In a nutshell, current quantum computers are still too fragile (in terms of noise and fidelity of quantum operations) to be used to perform full quantum algorithms like the celebrated Shor's factoring algorithm~\cite{Shor1994}.
However, recent results suggest that NISQ devices could be successfully used together with classical computers, to perform hybrid quantum/classical algorithms in which part of the computation (the most memory demanding) is performed on the quantum device, while the rest is performed on a classical machine.
In this way, the quantum overhead is reduced so that the necessary quantum operations can be performed, with a sufficient degree of accuracy, on current NISQ devices.
At the same time, it has been argued that the resulting architecture is still able to show a \textit{quantum advantage} over state-of-the-art implementations using purely classical algorithms~\cite{xiao2023}; though this last statement has been heavily debated~\cite{diez2021,bittel2021}.
Regardless of the presence or absence of quantum advantage, a common feature of NISQ devices is that they currently \textit{cannot} generate large amounts of entanglement.
This limitation comes hand-in-hand with noise: entangling quantum gates, like the CNOT, are quite noisy.
Therefore, the amount of CNOTs that can be used in a NISQ device is limited.
Hence, only area-law entanglement states can be reliably generated, while volume-law states are still out of reach. 

In parallel, the astonishing improvements in quantum simulator technologies, and the subsequent possibility of building mesoscopic quantum systems evolving in almost perfect isolation from their surrounding, have renewed the interest in understanding how isolated quantum systems thermalize.
Thermalization for isolated quantum many-body systems is currently understood in terms of the eigenstate thermalization hypothesis (ETH)~\cite{Deutsch1991,Srednicki1994,DAlessio2016}.
According to ETH, the energy eigenstates of \textit{generic} quantum many-body systems behave as the eigenvectors of random matrix theory (RMT) and, as such, they behave as effectively random vectors in the many-body Hilbert space.
Therefore, they display \textit{volume-law} entanglement for a generic bipartition of the system under investigation. 
There, an interesting research direction is in finding systems escaping ETH and thermalization.
Among the mechanisms proposed, we mention many-body localization~\cite{Huse2016} and \textit{many-body scarred} systems~\cite{bernien2017,turner2018a,turner2018b,shibata2020,iadecola2020,srivatsa2020,dooley2021,desaules2023,sanada2023}, the latter being the actual focus of this work.

A many-body scarred system is a quantum mechanical many-body system which \textit{generically} obeys ETH (with energy eigenstates displaying volume law entanglement) -- except for a few \emph{isolated} excited eigenstates that deviate from volume-law entanglement and show area-law entanglement \footnote{The definition of many-body scarred system can differ slightly in the literature. 
In the original paper~\onlinecite{turner2018a}, dealing with the famous PXP model, the term many-body scarred was coined to refer to the situation in which \textit{a set} of anomalous area-law eigenstates, almost \textit{equally spaced} in energy, was showing a huge overlap with a particular product state (the N\'eel state).
The evolution of the system, once prepared initially in the N\'eel state, shows huge \textit{revivals} and marked Loschmidt echoes despite being thermalizing.
In subsequent papers~\onlinecite{srivatsa2020,iadecola2020,sanada2023}, the definition of many-body scarred systems has been weakened, to include systems generically exhibiting ETH behavior but showing just a few (or even a single) anomalous eigenstates with area-law entanglement.
In this paper, we will stick to this weaker definition of a many-body scarred system.}.
Although initially found for systems with interactions being very fine-tuned~\cite{turner2018a, turner2018b}, it has been shown that more general systems can host many-body scars.
These include spin-\(1\) chains~\cite{schecter2019weak}, disordered systems~\cite{shibata2020onsager's, mondragon-shem2021fate}, and systems perturbed away from the fine-tuned point~\cite{bull2019systematic, bull2020quantum, michalidis2020stabilizing, lin2020slow, surace2021exact, kolb2023stability}. 

Given that these special eigenstates are sporadic -- and generically immersed in a sea of thermalizing eigenstates -- their existence is quite puzzling.
For example, for disordered systems, a Mott-like argument would suggest that the volume-law eigenstates, in which the scars are immersed, should hybridize with these states and ultimately destroy them.
However, several mechanisms explaining their existence, together with their stability, have been proposed (see the reviews~\cite{serbyn2021quantum, moudgalya2022,chandran2023}).
At the same time, these anomalous states are usually not easy to target and characterize.
They are often found by full exact diagonalization of the many-body Hamiltonian and by computing the entanglement entropy of the eigenstates for a given bipartition, although some algorithms based on a matrix product state (MPS) Ansatz have been recently proposed~\cite{moudgalya2020large, zhang2023}.

In this paper, we show that the main limitation of NISQ devices, \textit{i.e.} the inability of current quantum computers to produce large amounts of entanglement, can be turned to be \textit{a resource}, to directly and effectively hunt for quantum many-scarred eigenstates.
More in detail, we show that \textit{shallow variational quantum circuits}, \textit{i.e.} quantum circuits obtained by repeating just a few layers of parametric gates, can be successfully trained (via a hybrid quantum/classical variational scheme) to target the many-body scarred eigenstates in an otherwise thermalizing spectrum. 

The rationale behind this success is easy to imagine. Since the circuit cannot produce a large amount of entanglement, it cannot target any of the volume-law eigenstates present in the spectrum.
Therefore, a minimization algorithm trying to target an excited eigenstate is forced to either fail converging or to target just the sporadic scarred many-body eigenstates when present.
Our extensive numerical investigation will show that this simple logic works as expected, with shallow circuits being successful in targeting special area-law eigenstates.

As a consequence, the algorithm is \textit{robust}: many-body scarred eigenstates can be targeted by taking fairly generic quantum circuits allowing for a flexible circuit design.
Our Ansatz for the circuit does not assume any prior knowledge of the system under investigation beyond its symmetries, and therefore it can be used as a diagnostic tool for discovering many-body scarred eigenstates in new systems.
This robustness property represents the main strength of the algorithm.

\sect{Theoretical background}
Variational quantum algorithms (VQAs) have emerged as the leading strategy for achieving quantum advantage through today's NISQ devices. 
Any VQA can be split into three main features: an ansatz formed by a \textit{parametric} quantum circuit -- \textit{i.e.} a quantum circuit containing gates depending on continuous parameters (denoted collectively with \(\bth\)) --  to be run on a quantum computer;
a cost function (denoted with \(C(\bth)\)), whose minima encode the solution to the problem under investigation; and, finally, a classical optimization method which is used to target the minima of the cost function. 
Among VQAs, the variational quantum eigensolver (VQE) is used to approximate the ground state of a given quantum many-body Hamiltonian, \(H\)~\cite{tilly2022,kandala2017}.
To succeed in this task, the variational principle of quantum mechanics prompts the following cost function:
\begin{gather}
    \label{eq:ground_state_ansatz}
    C(\bth) = \mel{\psi(\bth)}{H}{\psi(\bth)} = \langle H \rangle ,
\end{gather}
where \(\ket{\psi(\bth)} = U(\bth)\ket{\psi}\) is the state produced by the parametric quantum circuit, and \(U(\bth)\) is the unitary operator representing the VQE ansatz.

By construction, the global minimum of \(C(\bm{\theta})\) provides the best approximation to the ground state of \(H\) -- among all the states that can be represented by the parametrized unitary circuit \(U(\bth)\).
This has led to the notions of expressivity/trainability trade-off and barren plateau. 
Indeed, it has been observed that very expressive circuits, \textit{i.e.} circuits with many variational parameters and that are capable of generating large portions of the full many-body Hilbert space, suffer severe drawbacks at the level of trainability. 
To be more precise, by increasing the number of variational parameters, the landscape described by the cost function \(C(\bth)\) tends to be rather flat, with many local minima and small values of the cost function gradient. 
This phenomenon has been dubbed barren plateau in the literature, and huge research has been devoted to finding ways and algorithms to reduce this optimization difficulty~\cite{mcclean2018, sim2019expressibility, kim2021universal, cerezo2021,marrero2021,patti2021,wang2021,uvarov2021,holmes2022,sack2022, kim2022quantum, zhang2022, ge2022the, larocca2023theory, matos2023characterization, heyraud2023efficient, kim2023quantum, incudini2023resource}. 

In this work, we aim to target scarred eigenstates, which are highly excited states, though they obey an area-law entanglement and thus they can be generated with a fairly reduced amount of variational parameters and entangling gates.
Therefore, the cost function in Eq.~\eqref{eq:ground_state_ansatz} must be modified, to target states different from the ground state.
While there are proposals for extending VQE to excited states \cite{nakanishi2019,zhang2021,liu2023}, this is considered a significantly more challenging task than ground-state computation.

Here we propose a VQE algorithm for detecting scar states, which we nickname VQE-S.
It is based on the following cost function:
\begin{gather}
    \label{eq:cost}
    C(\bth) = a \,\langle (H-E)^2\rangle + b \,(\langle H ^2\rangle - \langle H\rangle^2)+ c\, f_{\mathrm{symm}},
\end{gather}
where \(E\) is the target energy (\textit{i.e.} the energy value around which we aim to find a scarred eigenstate, if existent).
The first term in Eq.~\eqref{eq:cost} ensures that the variational state \(\ket{\psi(\bth)}\) has a mean energy as close as possible to \(E\);
while the second term gets minimized for \(\ket{\psi(\bth)}\) being an eigenstate.  
The function \(f_{\mathrm{symm}}\) is an additional penalizing term, whose explicit form is model-dependent, and that in models having different symmetry sectors can be used to target eigenstates living in a specific sector~\cite{turner2018a,iadecola2020,srivatsa2020, park2021efficient}.

The cost function~\eqref{eq:cost} defines a multi-objective (Pareto) optimization problem, with more than one objective function to be optimized simultaneously.
The real and positive parameters \(a\), \(b\), and \(c\), with the constraint \(a+b+c=1\), control the relative strength of the three terms in the total cost function.
Differently from previous attempts~\cite{gustafson2023}, the cost function in Eq.~\eqref{eq:cost} is \textit{model-agnostic}, \textit{i.e.} it does not assume any prior knowledge of the model under investigation, beyond its symmetries.
In particular, it does not assume a knowledge of the presence of scarred states in the spectrum.

The second point in VQAs is the circuit ansatz, which often represents a critical choice for the success of the algorithm. 
However, we have deliberately decided to consider fully agnostic ansatz only, to demonstrate the generality of our algorithm. 
In detail, we have considered three ansatze -- already heavily discussed in the literature and for which the trade-off between trainability and expressivity has been extensively studied~\cite{wu2021,bravo2020,sim2019} -- which we describe in the Supplemental Material (SM). %Methods. 
The results with these three ansatze are generically good, though they should be considered ``plug-and-play'' since there is no theoretical argument predicting which of the three performs %behaves 
best with a given model.
In the following, the best result for each Hamiltonian will be shown (see SM for a comparison).

\begin{figure}[h!]
	\includegraphics[width=8.0cm]{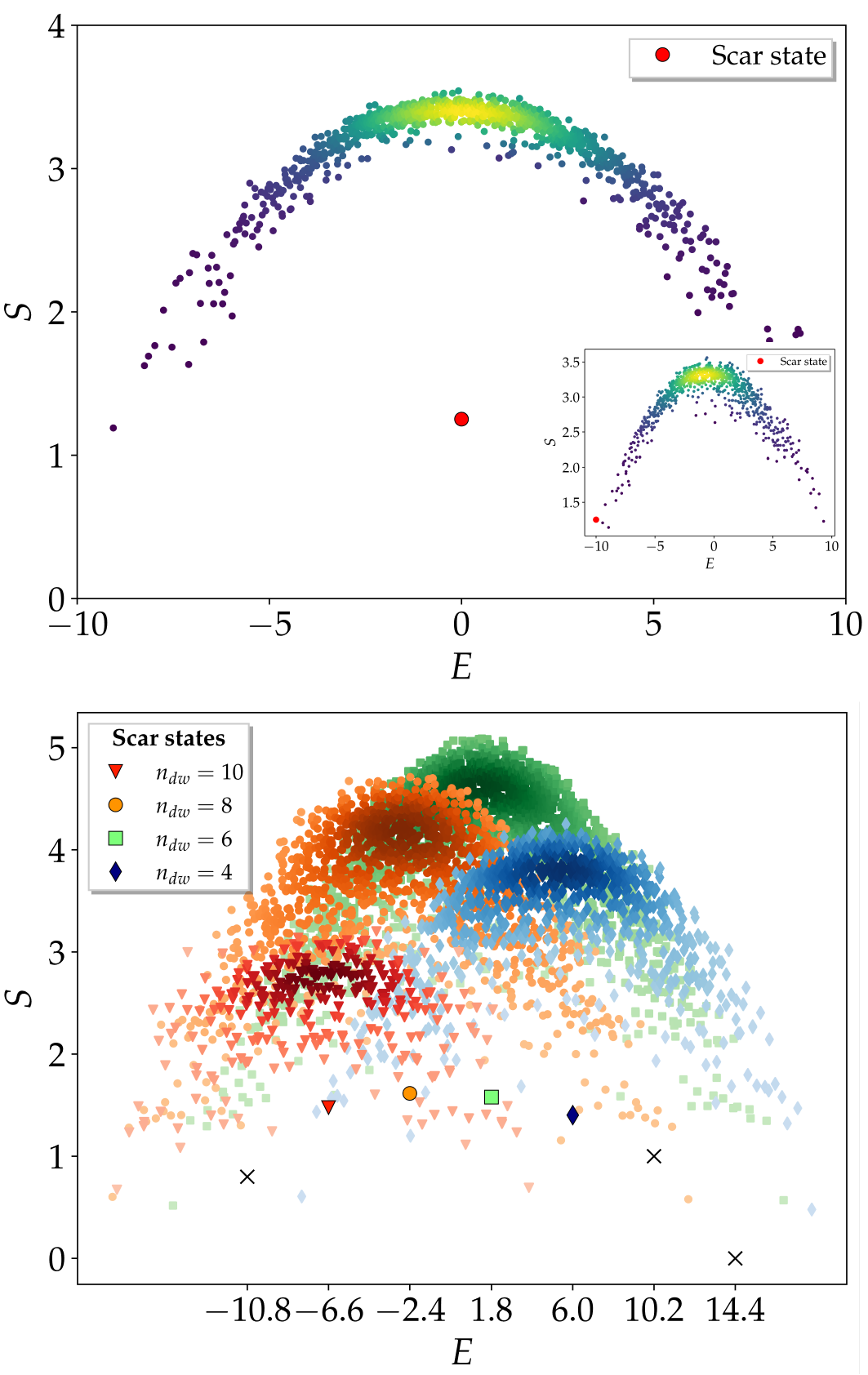}
	\caption{
        Top: Density plot of eigenstates spectrum of the \(N/2\) bosons subspace of \(H_1\) with \(N=12\), the scar is highlighted in red.
        In the inset the model parameters are tuned to make it scarless.
        Bottom: Density plot of the eigenstates of \(H_2\) with \(N=14\).
        Shown here are only the states for the 
    %$(\sigma^z_1,\sigma^z_N)=(+1,+1)$ 
        \(\ket{0\dots0}\) subspace and \(n_{dw}=4\) (blue diamonds), \(n_{dw}=6\) (green squares), \(n_{dw}=8\) (orange circles), and \(n_{dw}=10\) (red triangles).
        The darker colors indicate regions with higher densities of eigenstates.
        The scar states are highlighted according to the legend.
        The crosses represent scars of the other subspaces.
    }
    \label{fig:spectra}
\end{figure}

\sect{Numerical simulations}
To illustrate the effectiveness of our variational quantum algorithm in hunting for scar states, we consider two models ordered by increasing complexity:
the first contains a single scar, while the second contains towers of scars (see SM for details on the Hamiltonians investigated).
We first consider the following 1D Hamiltonian of \(N\) hard-core bosons placed in a circular lattice~\cite{srivatsa2020}: %$z_j=e^{i 2 \pi j/N}$ %srivasta2020
\begin{equation}
    \begin{array}{c}
        H_1=\sum_{i\neq j} G_{ij}^{A} d_i^\dagger d_j+\sum_{i\neq j} G_{ij}^{B}  n_i n_j
        \\\\
        +\sum_{i\neq j\neq l} G_{ijl}^{C} d_i^\dagger d_l n_j
        +\sum_{i} G_{i}^{D}  n_i +G^E,
    \end{array}
    \label{eq:model1H}
\end{equation}
where the operators \(d_j\), \(d_j^\dagger\), and \(n_j = d^\dagger_j d_j\), acting on the site \(j\), are the annihilation, creation, and number operators for hard-core bosons, respectively; while \(G\)s are constants.
This Hamiltonian is particularly suitable since it has only one scar state in the middle of the spectrum at \(E=0\) [Fig.~\ref{fig:spectra} (top panel)],
and also the location of the scar can be adjusted by tuning a parameter, thus the algorithm can be compared with basically the same model with the scar removed [see Fig.~\ref{fig:spectra} (inset of the top panel),
where the scar state of the original model has been moved out of the energy spectrum with ergodic eigenstates].
\(H_1\) has one conserved quantity which is the total number of bosons operator, defined as \(\hat{N}_{b}=\sum_i n_i\), associated with the quantum number \(n_b\).
The scar state is located in the \(n_b=N/2\) symmetry sector; therefore, the symmetry term in Eq.~\eqref{eq:cost} is \(f_{symm}=\langle (\hat{N}_b-n_b)^2\rangle\).\\

To further validate the algorithm we also consider the following spin-1/2 model on a \(N\)-length chain whose Hamiltonian reads~\cite{iadecola2020}:
\begin{gather}
    H_2=\lambda \sum_{i=2}^{N-1} (\sigma^x_i-\sigma^z_{i-1}\sigma^x_i \sigma^z_{i+1} ) + \Delta \sum_{i=1}^N \sigma^z_i + J\sum_{i=1}^{N-1} \sigma^z_{i} \sigma^z_{i+1}.
\end{gather}
The Hamiltonian has a spatial inversion symmetry with eigenvalues \(\pm1\) associated to the operator \(\hat{I}=\prod_{i=1}^N\sigma^x_i\) and it also conserves the number \(n_{dw}\) of Ising domain walls, associated to the operator \(\hat{N}_{dw}=\sum_{i=1}^{N-1}(1-\sigma^z_i\sigma^z_{i+1})/2\). 
This model shows two towers of scar eigenstates, related by the inversion of all spins. 
Each scar is associated with a definite number of domain walls, and the scars of the same tower have alternating inversion quantum number.
Since \([H_2,\sigma^z_{1,L}]=0\) the edge spins are constants of motion and can be fixed;
each tower is associated with edges \(\ket{0\dots0}\) and \(\ket{1\dots 1}\).
The tower for edges \(\ket{0\dots 0}\) is shown in Fig.~\ref{fig:spectra} (bottom panel) (crosses and highlighted symbols). 
In this figure only the states with \(n_{dw}=4,6,8,10\) for \(N=14\) are plotted, since in the following we will target only these states.
For this Hamiltonian, the symmetry term in Eq.~\eqref{eq:cost} is \(f_{symm}=\langle (\hat{N}_{dw}-n_{dw})^2\rangle\).

\begin{figure}[h!]
    \includegraphics[width=8.0 cm]{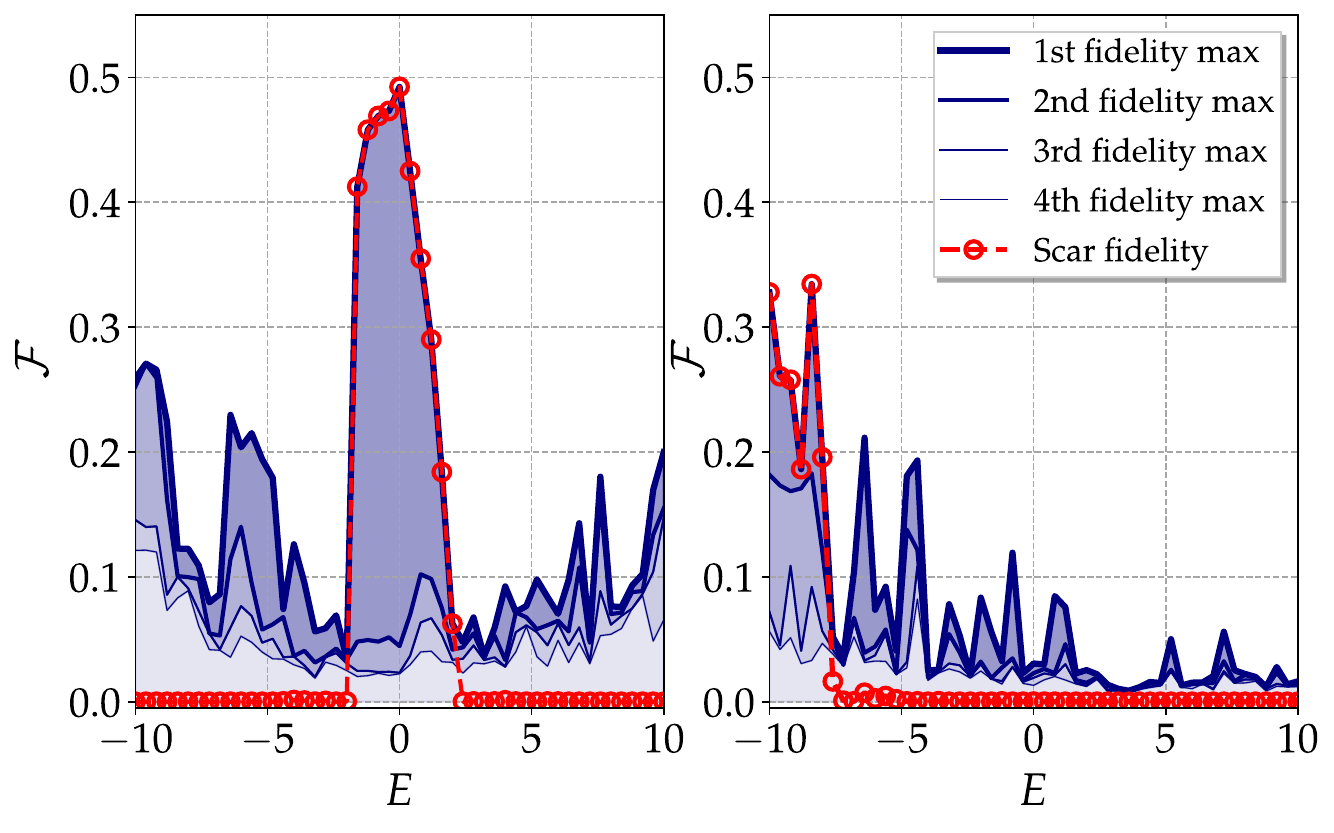}
	\caption{
        Comparison of the four largest fidelities, as a function of target energy, between the VQE-S state and the eigenstates of \(H_1\) with a scar (left) and scarless (right).
    }
	\label{fig:mod1energyScanComparison}
\end{figure}

To show that our VQE-S algorithm can detect the presence of quantum many-body scars, we repeat the algorithm for various target energies. 
In the cost function, we have taken the values \(a=0.05\), \(b=0.25\), \(c=0.7\).
The algorithm is very robust, since there is a broad range of values for which convergence is good (see SM).
First, we consider Hamiltonian \(H_1\).
In Fig.~\ref{fig:mod1energyScanComparison} (left panel) the four highest fidelities  \(\mathcal{F}_i=|\braket{\psi_{VQE}}{\phi_i}|^2\) between the optimized VQE-S state \(\ket{\psi_{VQE}}\) and the eigenstates of the Hamiltonian \(\ket{\phi_i}\) are plotted as a function of the target energy \(E\).
The results are compared with a modified version of \(H_1\) without the scar state (right panel). 
In contrast to the model without scars, a peak in fidelity due to the scar state is evident in the scarred model.

\begin{figure}[h!]
	\includegraphics[width=8.0 cm]{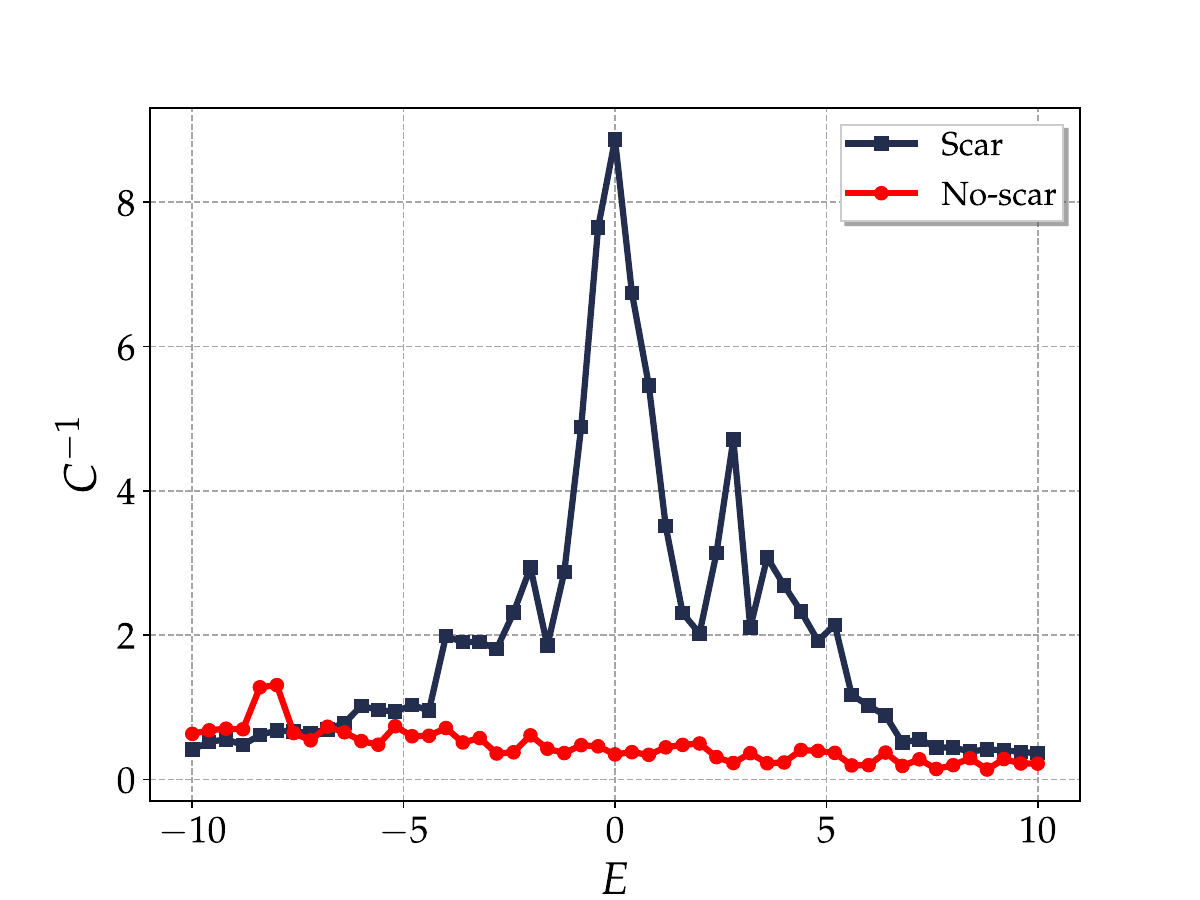}
	\caption{
        Comparison of the inverse of the cost function as a function of the target energy of the VQE-S state, for Hamiltonian \(H_1\) with scar (solid lines) and scarless (dashed lines).
    }
	\label{fig:mod1energyScanCostComparison}
\end{figure}

Fidelities, however, are not easily accessible experimental quantities. 
Indeed, they would require a complete knowledge of the Hamiltonian spectrum.
Therefore, to search for scar states unknown a priori, we examined the inverse of the cost function in Fig.~\ref{fig:mod1energyScanCostComparison}.
This quantity has a clear peak at the energy of the scar (similar to that obtained for fidelity), whereas for the model without a scar, there are no prominent features. 
The cost function can then be used as a convenient probe to discover the presence of scar states. 
A key point to be stressed is that the depth ({\it i.e.} the number of steps) of the ansatz must be kept shallow. Indeed, this has a twofold role: it avoids barren plateaus, and it also limits the amount of generated entanglement, thus exploiting the intrinsic properties of NISQ. 

\begin{figure}[]
    \includegraphics[width=8.0 cm]{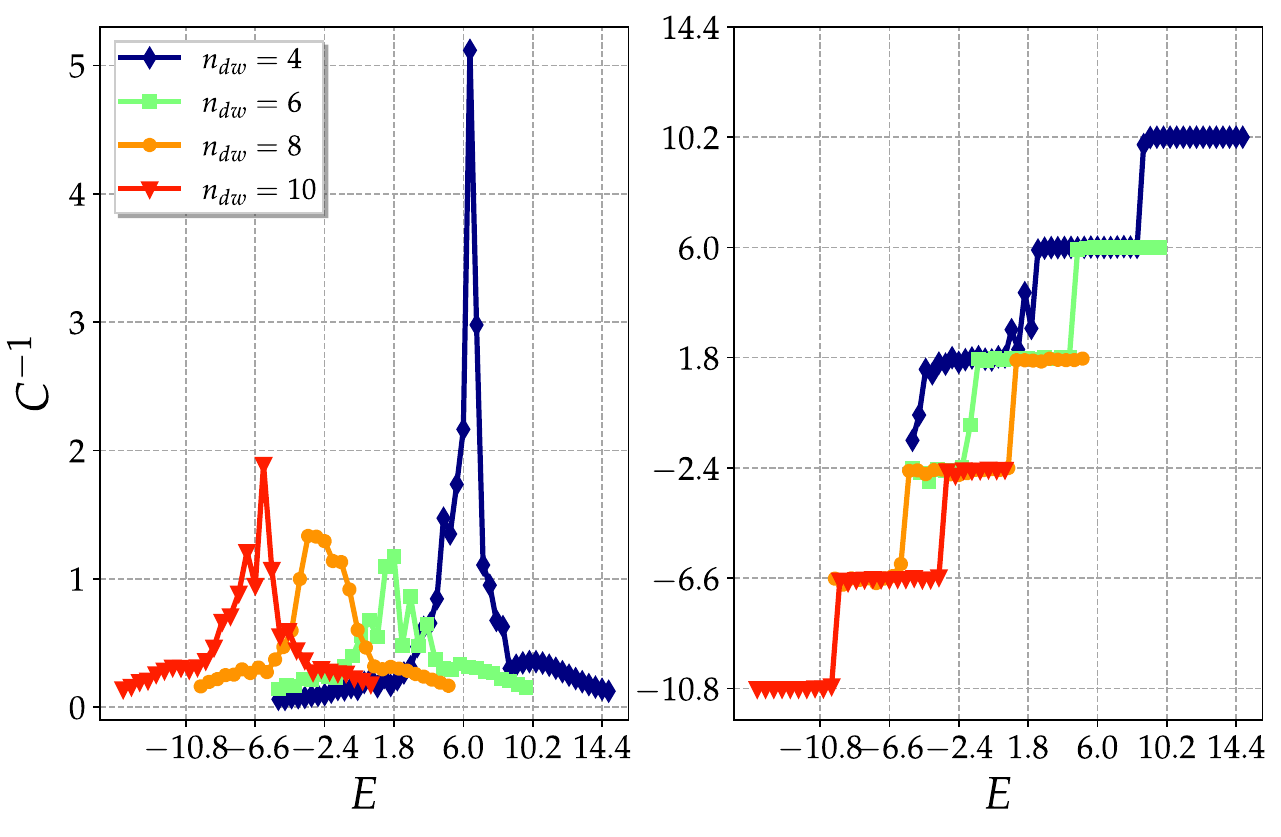}
	\caption{
        Inverse of the cost function (left) and mean energy (right) for Hamiltonian $H_2$, with $N=14$ spins, and parameters values $\lambda = 1$, $\Delta = 0.1$, and $J = 1$.  
    }
	\label{fig:mod2energyScanCostComparison}
\end{figure} 

We have repeated the same analysis for Hamiltonian \(H_2\), which exhibits towers of scars instead of just one.
In this case, we have targeted the most densely populated subspaces, i.e., \(n_{dw}=4,6,8\), and \(10\) for \(N=14\) spins.
Notice that, since \([H_2,\sigma^z_{1,N}]=0\), the edge spins are constants of motion and can be fixed, and thus the ansatz is applied at the bulk of the spin chain only. 
Fig.~\ref{fig:mod2energyScanCostComparison} (left) shows a clear peak in the inverse cost function at the energy of the scar state for the selected value of \(n_{dw}\).
Even more striking is the behavior of the mean energy \(\langle H\rangle\) as a function of the target energy, shown in Fig.~\ref{fig:mod2energyScanCostComparison} (right). 
A clear plateau sequence can be seen at the energies of the scar states, even for scar states corresponding to values of the quantum number \(n_{dw}\) other than those selected in the cost function.

\sect{Discussion}
NISQ devices are unable to generate a large amount of multipartite entanglement, and therefore generic volume law quantum many-body states that obey the ETH hypothesis cannot be accessed.
In this work, we have shown that this limitation can be turned into a resource, to hunt, via shallow variational quantum circuits, for many-body scar states.
Perhaps surprisingly, VQA converges to the scar states even though these are sporadic eigenstates immersed in a sea of unaccessible ETH eigenstates.

In comparison with the classical MPS approach~\cite{zhang2023}, we can say that the variational approach developed in this paper is particularly versatile. 
In the MPS approach, given a certain target energy, the algorithm tries to find the corresponding eigenstate among the states satisfying the MPS ansatz. 
The procedure requires a fine-tuning of the target energy and an initial state with a non-zero overlap with the target scar state. 
While the second condition is rather generic -- two, randomly taken, states do likely have a certain amount of overlap -- the first condition is quite strong and it does often require prior knowledge of the system under investigation or some feedback mechanisms to adjust the target energy.
Moreover, the choice of the bond dimension of the MPS ansatz adds an extra degree of arbitrariness in the algorithm, and it can lead to disparate outcomes. 
As a result, the fidelities obtained with the MPS method are in general very high, at the expense of algorithmic simplicity. 
Moreover, the MPS approach (to our knowledge) does not have an easy generalization to 2D and 3D many-body systems, while our variational hybrid classical/quantum algorithm does not suffer from these limitations.

Our approach could also be used to investigate the presence of scar states when approaching the classical limit.
In the latter case, scars are eigenstates of a classically chaotic quantum system with higher probability density around the paths of classical unstable periodic orbits.
Such states introduce corrections to the universal spectral statistics of the random matrix theory.
First observed in classically chaotic quantum billiards~\cite{McDonald1979,Heller1984}, such scar states also feature in chaotic quantum maps~\cite{Faure2003}, easily amenable to efficient quantum simulations~\cite{Benenti2001}.
Therefore our VQE-S method could be conveniently used to investigate scar states in the quantum to classical transition, a task that does not seem easy to address with the MPS approach.

\begin{acknowledgments}
    We would like to thank Dogyun Ko, for discussions and collaborations on related subjects.
    GC and GB acknowledge financial support from PRIN MUR (Grant No. 2022XK5CPX) and from INFN through the project “QUANTUM”.
    AA and DR acknowledge financial support from the Institute for Basic Science (IBS) in the Republic of Korea through the project IBS-R024-D1.
    DR thanks FAPESP, for the ICTP-SAIFR grant 2021/14335-0 and for the Young Investigator grant 2023/11832-9, and the Simons Foundation for the Targeted Grant to ICTP-SAIFR.
    MC acknowledges financial support from PRIN MUR (Grant No. 2022PH852L).
\end{acknowledgments}

%%%%%%%%%%%%%%%%%%%%%%%%%%%%%%%%%%%%%%%%%%%%%%%%%%%%%%%%%%%%%%%%

\bibliography{biblio.bib}

\clearpage

\section*{Supplemental Material}

In this section, we provide further details for the Hamiltonian models studied in the main text, as well as details of our numerical simulations and additional data to support the robustness of 
the VQE-S algorithm for finding scar states.

\subsection{Details of the studied models}

The coefficients of Hamiltonian $H_1$ are given by
\begin{align}
    G_{ij}^A&=-2 w_{ij}^2,\notag\\
    G_{ij}^B&=(2-\alpha)w_{ij}^2 + 4\sum_{l (\neq j\neq i)} w_{ij}w_{il},\notag\\
    G_{ijl}^C&=-\alpha w_{ji}w_{jl}  ,\notag\\
    G_{i}^D&= -2 \sum_{j(\neq i)} w_{ij}^2 - \sum_{j\neq l (\neq i)} w_{ij} w_{il},\notag\\
    G^E&= \frac{-N(N-2)(N-4)}{6} ,
\end{align}
where $w_{ij} =\dfrac{z_i+z_j}{z_i-z_j}$, with $z_j=e^{i 2 \pi j/N}$, and the real parameter $\alpha$ can be selected to shift the energy of the scar state and thus to move it in the spectrum of $H_1$.
The Hamiltonian is thus a real operator with some non-local terms.
As written in the main text, the particle number, associated with the operator $\sum_i n_i$, is a conserved quantity and, in order to construct a model with a thermal spectrum,
it is possible to add a small amount of random disorder to the lattice site positions to avoid any additional symmetries: 
\begin{equation}
    z_j= e^{i 2\pi (j+\gamma_j)/N},\quad j \in \{1,2,\dots,N\},
\end{equation}
where $\gamma_j$ is a random number chosen with constant probability density in the interval $\big[ -\frac{\delta}{2},\frac{\delta}{2}\big]$ where $\delta$ is the disorder strength. 
We call $\ket{\Psi_{scar}}$ the eigenstate of $H_1$ that has all the properties of a quantum many-body scar state.
In the simulations shown in the main text, we used the disorder model fixing $\delta=0.5$ and we chose $\alpha=-2.5$, to put $\ket{\Psi_{scar}}$ in the middle of the thermalizing spectrum, for $N=12$:
$$
    \ket{\Psi_{scar}}= \mathcal{N}\sum_{n_1,n_2,\dots,n_N} (-1)^{\sum_j(j-1)n_j}\delta_n 
$$
\begin{equation}    
    \times \prod_{i<j}(z_i-z_j)^{2n_i n_j - n_i - n_j}\ket{n_1,n_2,\dots,n_N}.
\end{equation}
Above $n_i \in \{ 0,1\}$ is the number of hardcore bosons in the $i$-th lattice site and $\delta_n=1$ if $\sum_i n_i=N/2$ ($\delta_n=0$ otherwise), thus $\ket{\Psi_{scar}}$ is also an eigenstate of the bosons number operator.
Hardcore bosons operators can be easily converted into Pauli spin operators and thus, the Hamiltonian $H_1$ can be effortlessly measured on a quantum computer by conveniently rotating the measurement axis.
In the main text, we compared the results of this model with a "scarless" one, in which we shifted the position of the scar by fixing $\alpha=0$.

Hamiltonian $H_2$ presents two towers of scar eigenstates, eigenstates of the number of domain wall operator.
The first tower is given by
$$
    \ket{\Psi^{(k)}_{scar}}=\frac{1}{k!}\binom{N-k-1}{k}^{-1/2} 
$$
\begin{equation}
    \times \Bigg[ \sum_{i=2}^{N-1}(-1)^i P_{i-1}\sigma^{+}_i P_{i+1} \bigg]^k \ket{0,\dots,0},
    \label{eq:scarstatesMod2}
\end{equation}
where $P_j=(1+\sigma^z_j)/2$, $\sigma^{\pm}_j= (\sigma^x_j\pm i\sigma^y_j)/2$ and $k\in\{0,\dots ,N/2-1\}$.
The second tower is simply obtained by flipping the spins, i.e. $\ket{\Phi^{(k)}_{scar}}=\hat{I}\ket{\Psi^{(k)}_{scar}}$. 
In the main text, we set $\lambda=1$, $\delta=0,1$ and $J=1$. For generic values of $\delta$ and $J$, the model is not-integrable; adding a finite $\delta$ introduces a highly non-local interaction term, breaking integrability. Physically, the state $\ket{\Psi^{(k)}_{scar}}$ contains $k$ magnons (i.e., spin flips), each carrying momentum $\kappa = \pi$. The states defined in Eq.~\ref{eq:scarstatesMod2} generically have sub-volume-law entanglement entropy, thus demonstrating explicitly their ETH-violating nature.

\subsection{Comparison of the ansatz used}

In this subsection, we provide evidence that the specific choice of the circuit does not affect the robustness of our findings. 

\begin{widetext}
\begin{figure*}[t]
    \includegraphics[width=\textwidth]{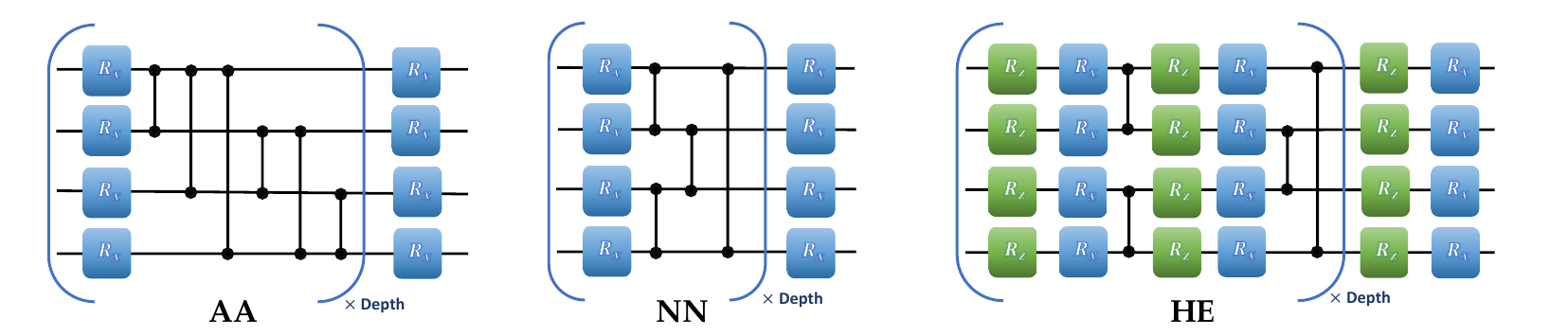}
	\caption{
        Ansatz circuits used in this work. From left to right:
        All-to-all entanglement ansatz (AA), 
        Nearest-Neighbors entanglement ansatz (NN), 
        Hardware Efficient ansatz (HE).
    }
    \label{fig:ansatz}
\end{figure*}
\end{widetext}

The ansatz used in this work -- that we dub All-to-All entanglement ansatz (AA), Nearest-Neighbors entanglement ansatz (NN), and Hardware Efficient ansatz (HE) -- are pictured in Fig.~\ref{fig:ansatz}.
In general, the depth of the circuits was kept shallow to avoid barren plateau and to limit the entanglement generated, and thus favor the search for area-law scar states.
These ansatz have been used extensively in many VQA, and they are the standard for NISQ systems~\cite{wu2021,bravo2020,sim2019}.
In this work, the entangling gate used is the controlled-Z gate.
% As it was stated before, the three ansatz of Fig.~\ref{fig:ansatz} and similar variations on the theme (it is possible to switch the axis of single qubit gates and entangling gates) should be considered plug-and-play, as there is no theoretical argument that can suggest which of the three behaves best with a given model. 

\begin{figure}[h!]
    \includegraphics[width=8.0 cm]{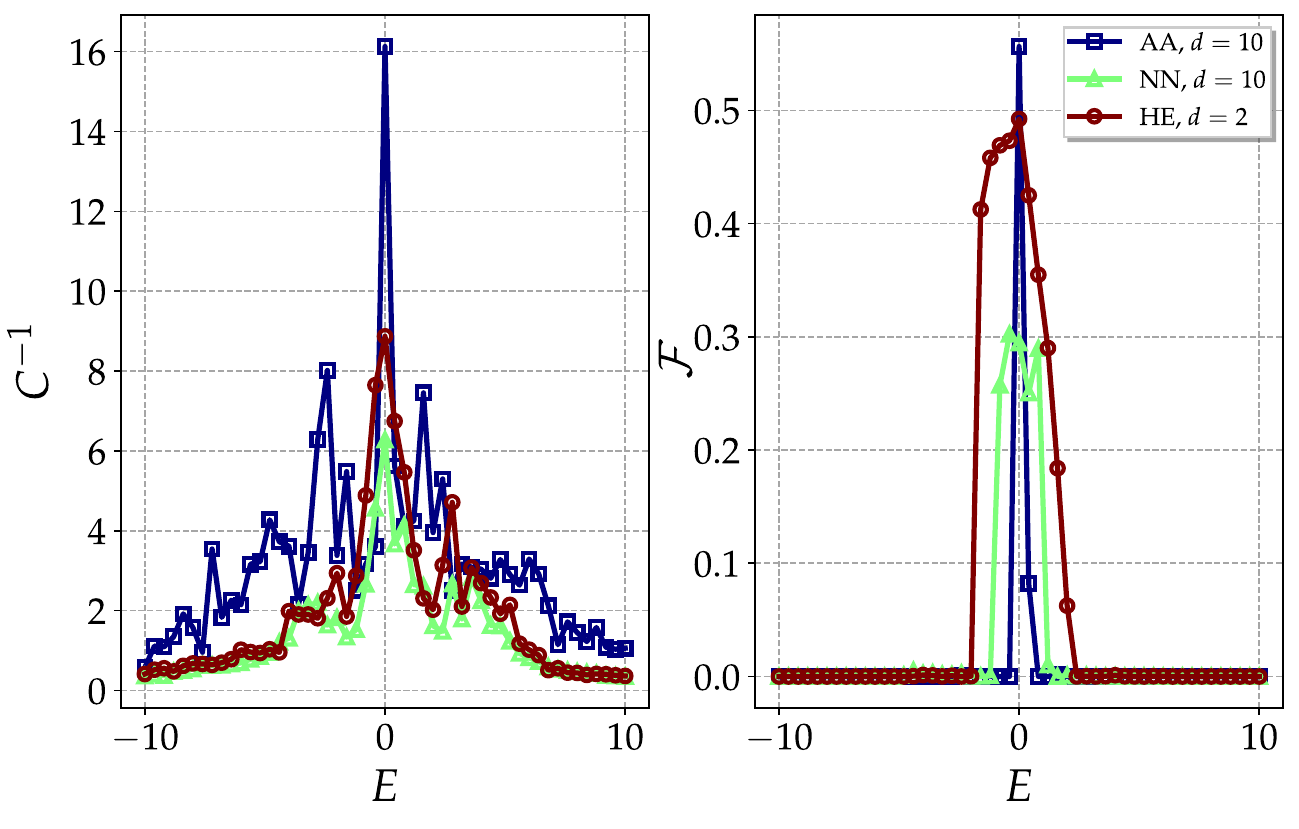}
	\caption{
        Comparison of the cost function inverse (left) and fidelity with the scar state (right) as a function of the target energy, for the ansatz considered in this and Hamiltonian $H_1$.
        Here we set $\text{depth}=10$ for AA and NN ansatz, and $\text{depth}=2$ for the HE ansatz.
        Cost function hyper-parameters are $a=0.05$, $b=0.25$ and $c=0.70$.
        For each target energy, we performed 5000 iterations for AA and NN and 1000 for HE.
    }
    \label{fig:ansatzComp}
\end{figure}

In what follows, We compare these three ansatz for $H_1$, running simulations for different target energies.
For the AA and the NN ansatz, we set a depth of $d=10$, and the number of optimization iterations is fixed at 5000, while for the HE ansatz we set a depth of 2 (to have a comparable number of parameters between NN, AA and HE) and the number iterations is taken to be 1000, as in the main text.
All other parameters were kept fixed, as before.
Figure~\ref{fig:ansatzComp} shows the inverse cost function and the fidelity with the scar state as a function of the target energy for the three ansatz.
In all three cases there is clearly a peak centered at $E=0$.
The peak in the AA case, despite being higher both for cost function and fidelity, is more ``noisy''.
The results are comparable in all three cases, however, the real advantage of HE ansatz is the number of iterations required to converge.
Therefore, we can conclude that for this model the HE ansatz is more convenient than the other two ansatz. 

On the other hand, for model $H_2$ we found more convenient either the NN or the AA ansatz. 
To summarize, for the data reported in the main text we used ansatz HE, $\text{depth}=2$ and $\text{iterations}=1000$ for Hamiltonian $H_1$, while for Hamiltonian $H_2$ we used the AA ansatz with $\text{depth}=5$ for $n_{dw}=4,6$ and the NN ansatz with $\text{depth}=10$ for $n_{dw}=8,10$, in both cases with $\text{iterations}=1000$. 

The main point to stress here is that, as anticipated, these differences are all quantitative, while the qualitative behavior is rather robust and unchanged when changing the circuit ansatz.

\subsection{Simulation details}

All the simulations have been performed using Python, in particular we acknowledge the use of Qulacs~\cite{qulacs} for fast quantum circuits and state vector simulations.
As classical optimizer we used ADAM (short for Adaptive Moment Estimation) which is an optimization algorithm commonly used in training deep neural networks.
It is an extension of stochastic gradient descent that is based on adaptive estimation of first-order and second-order learning moments. 
The parameters (the rotation angles $\theta_i$ appearing in the parametric gates) were initially chosen randomly in a small range $[-\epsilon,\epsilon]$, with  $\epsilon\sim 10^{-2})$.

\begin{widetext}
\begin{figure*}[ht]
	\includegraphics[width=\textwidth]{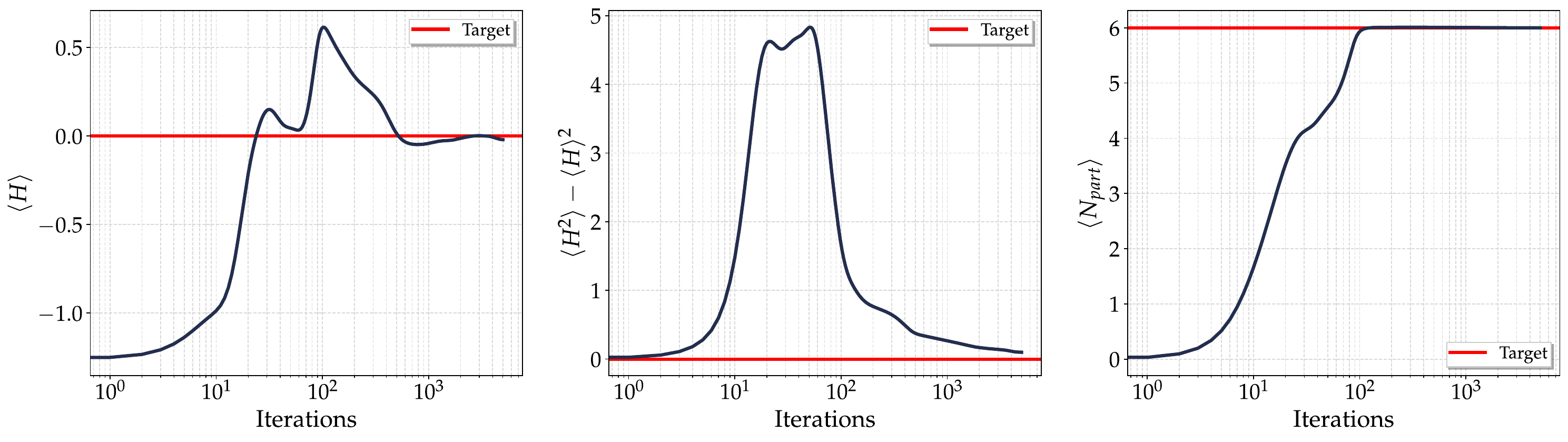}
	\caption{
        Mean energy, variance, and mean value of the number of particles for the VQE state as a function of the number of optimization iterations. 
        Red lines refer to the target values.
	}
    \label{fig:SUPconvergenceMOD1}
\end{figure*}
\end{widetext}

\begin{figure}[h!]
	\includegraphics[width=8.0cm]{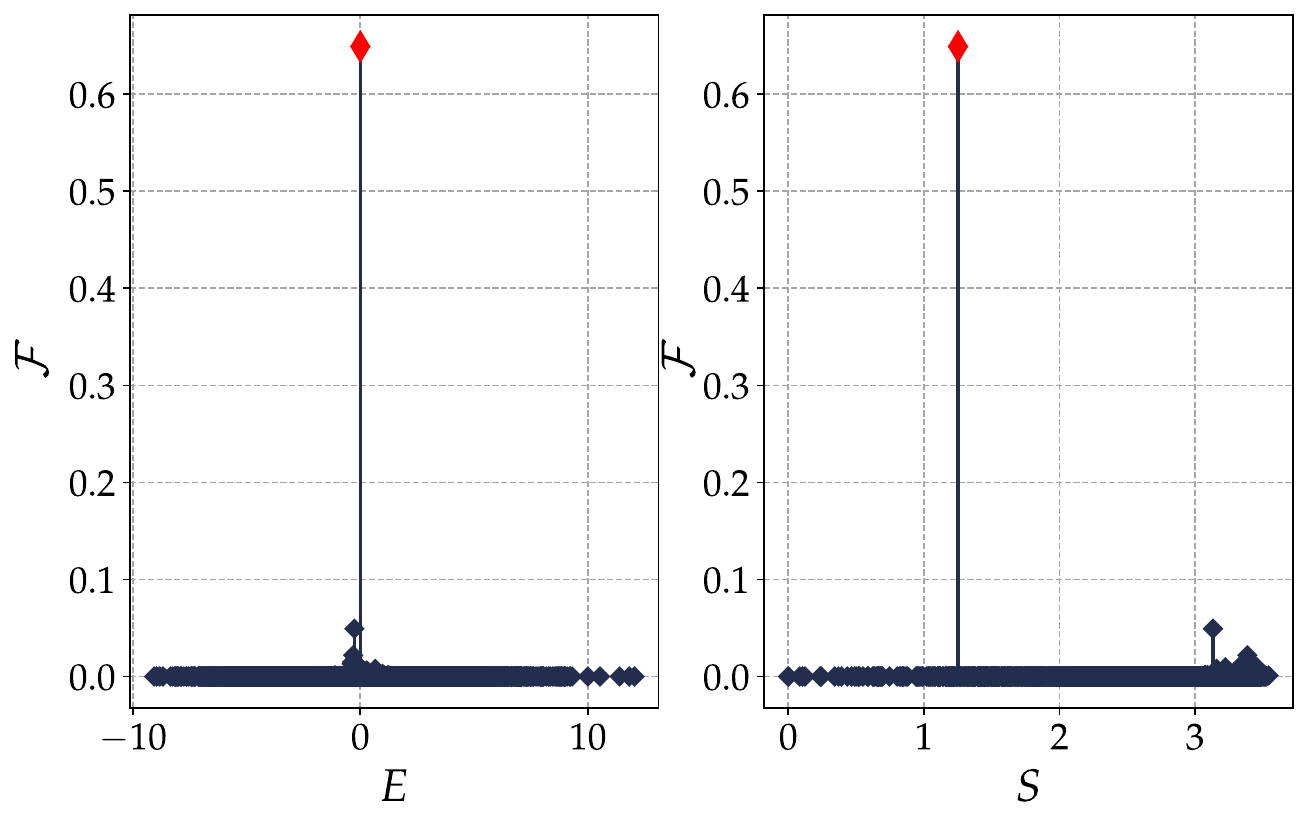}
	\caption{
        Fidelity of the VQE state with all the eigenstates of model $H_1$ as a function of their energy and entropy of entanglement.
        The red diamond denotes the scar state. 
    }
	\label{fig:SUPfidelityMOD1}
\end{figure}

Hereafter, we will focus on the $H_1$ scar for simplicity.
As a preliminary figure of merit to verify the convergence of our algorithm, we consider the fidelity of the VQE-S state $\ket{\psi_{\text{VQE}}}=U(\bm{\theta})\ket{\psi_0}$ and the scar state:
\begin{equation}
    \mathcal{F}=|\bra{\psi_{\text{VQE}}}\ket{\Psi_{scar}}|^2.
\end{equation}
For Hamiltonian $H_1$ the symmetry term in the cost function is $f_{symm}=\langle (\hat{N}_{part}-n_{part})^2\rangle$ where $\hat{N}_{part}$ is the hardcore bosons number operators and $n_{part}$ is the target state number of particles, which is set equal to $N/2$.
The target energy is set to $E=0$ in order to match the energy of $\ket{\psi_{scar}}$.
We tested the algorithm for $N=12$, the parameters of the cost function are arbitrarily fixed at $a=0.05$, $b=0.25$ and $c=0.70$. 
As can be seen in Fig.~\ref{fig:SUPconvergenceMOD1} the parameters of the cost function converge to the desired values.
After \(5000\) iteration steps the algorithm is stopped and the spectrum of the obtained VQE-S state is depicted in Fig.~\ref{fig:SUPfidelityMOD1} (left panel), where we can that the state with the greatest overlap 
is by far the scar state. Fig.~\ref{fig:SUPfidelityMOD1} (right panel) shows eigenstates fidelities as a function of entanglement entropy $S=-\text{Tr}\rho\text{log}\rho$. Here $\rho$ is the density operator for half chain. 
Although the scar state is far from being the lowest in entanglement, the cost function is capable of selecting it correctly.

\begin{figure}[h!]
	\includegraphics[width=8.0cm]{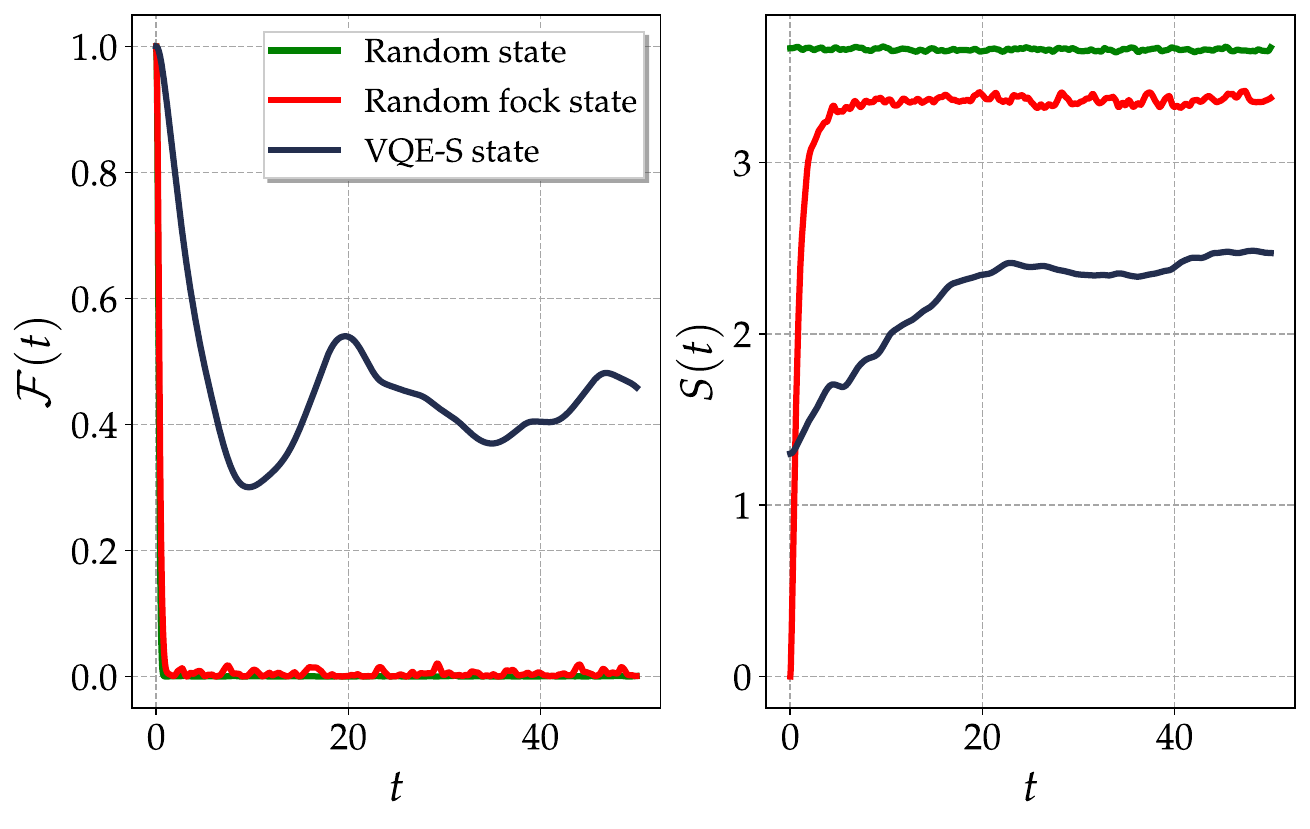}
	\caption{
        Dynamics of many-body fidelity (left) and the half-chain entanglement entropy (right) for various initial states at $N=12$. 
    }
	\label{fig:SUP_FSmod1}
\end{figure}

\begin{figure}[h!]
    \includegraphics[width=8 cm]{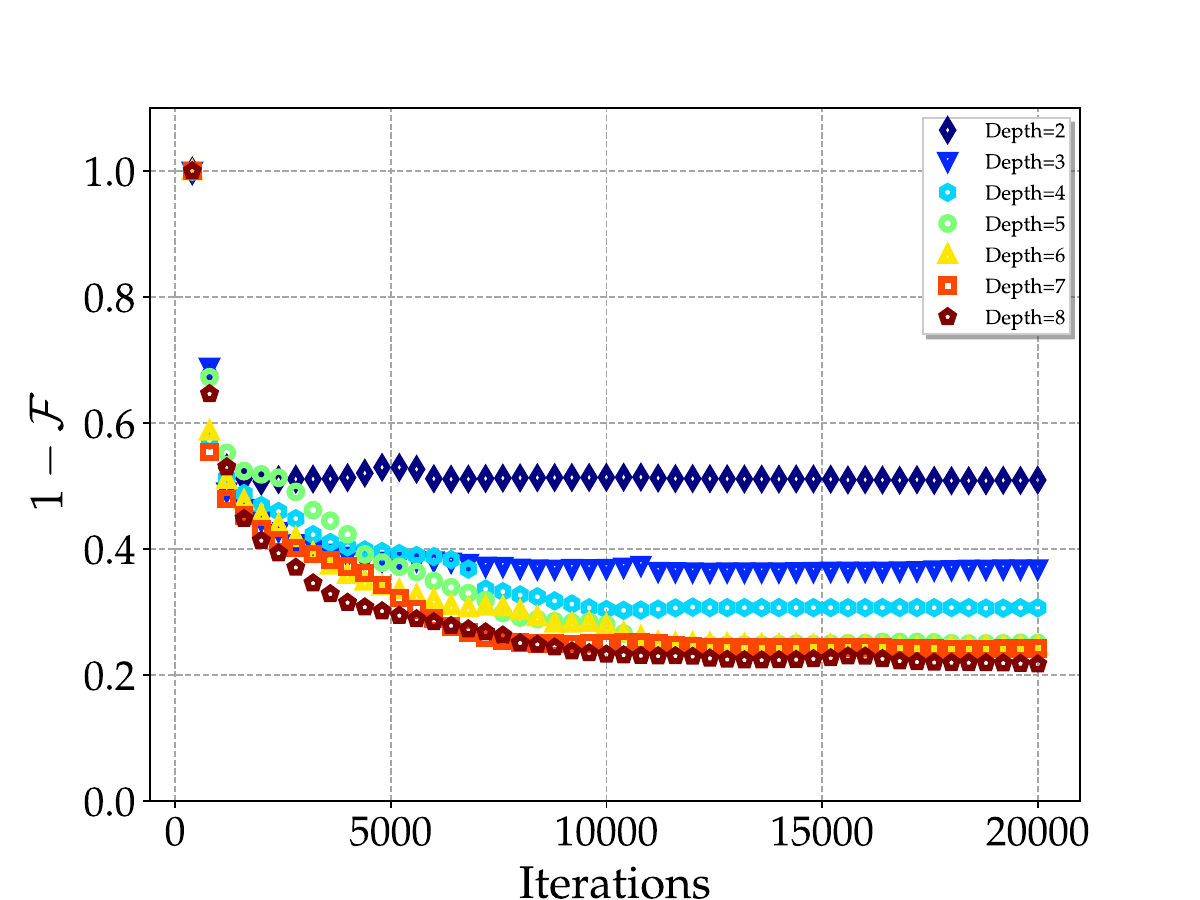}
    \includegraphics[width=8 cm]{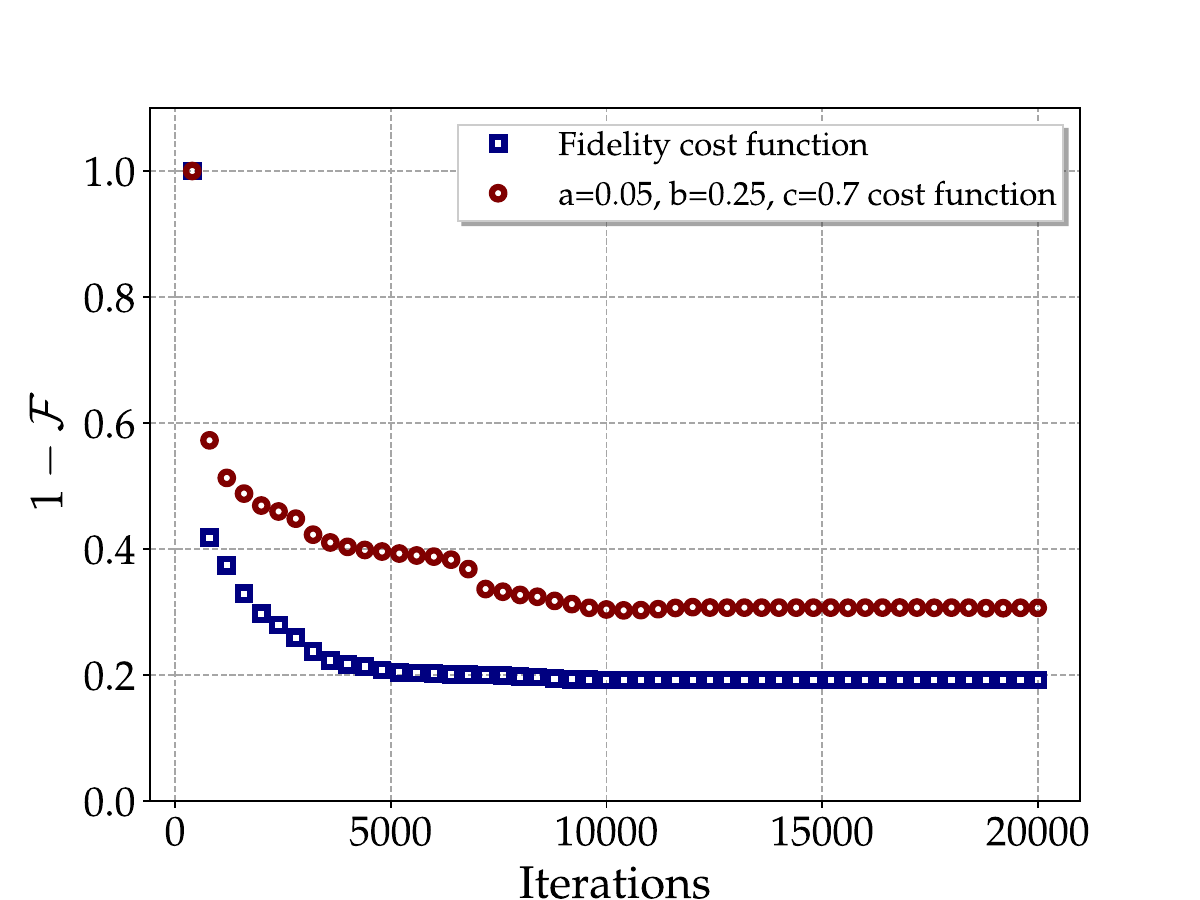}
	\caption{
        Infidelities with the scar state (top) as a function of the number of optimization steps, for various ansatz depths. Comparison (bottom) between the infidelity cost function and our 
        agnostic cost function with depth=4.
    }
	\label{fig:SUPscalingMod1}
\end{figure}

As has been amply demonstrated \cite{alhambra2020}, quantum states with finite overlap with scars exhibit long-lived oscillations (revivals) and thus non-thermal behavior and area-law entanglement entropies.
To check this feature for the states generated by the VQE, we calculated (see Fig.~\ref{fig:SUP_FSmod1}) the dynamical fidelity $\mathcal{F}(t)=|\bra{\psi_{\text{VQE}}(t)}\ket{\psi_{\text{VQE}}(0)}|^2$ and the half-chain dynamical entanglement entropy $S(t)=-\text{Tr}\rho(t)\text{log}\rho(t)$.
Here $\ket{\psi_{VQE}(t)}=e^{-iH_1t/\hbar}\ket{\psi_{VQE}(0)}$ and $\rho(t)$ is the half-chain density operator at time $t$, obtained after tracing the overall density operator $\ket{\psi_{VQE}(t)}\bra{\psi_{VQE}(t)}$ over the other half of the chain.
As is shown in Fig.~\ref{fig:SUP_FSmod1}, dynamical fidelity remains non-zero even for long times and dynamical entropies remain below the random state limit, indicating a state that does not thermalize.
These quantities are compared with those for random states and for a randomly chosen Fock state (i.e., a state $|i_1,...,i_N\rangle$, with $i_j$ randomly chosen to be either 0 or 1, for all $j=1,..., N$) within the same energy window.
In both cases, the fidelity rapidly decays to zero, and for a random Fock state the entanglement grows to near the value of the random state.
This check can be used as a first proof of the algorithm convergence, since it does not use any a priori knowledge of the Hamiltonian spectrum, nor the presence of a scar state.
%\textcolor{blue}{Should we mention something about this in the main text? and here, should we close the period with what we learned from this check?}
Finally, we have checked the working of our VQA as a function of the circuit depth and the number of optimization steps. 
As we can see from the infidelity ($1-\mathcal{F}$), as expected the convergence improves both by increasing the depth and increasing the number of optimization steps (top panel of Fig.~\ref{fig:SUPscalingMod1}).
In the lower panel of the same figure, we compare the result of the cost function used in our work with the best possible cost function, which is the infidelity between the VQE state and the scar state,
\begin{equation}
    C_\mathcal{F}(\bm{\theta})=1-|\bra{\psi_{VQE}(\bm{\theta})}\ket{\Psi_{scar}}|^2.
    \label{eq:infidelity_cost}
\end{equation}
%(see Eq. \ref{eq:infidelity_cost}),
This latter cost function, however, requires prior knowledge of the scar state.
As can be seen, our agnostic cost function, which instead assumes no a priori knowledge (not even about the existence of scars), exhibits, asymptotically in the number of iterations, a performance comparable to that obtained using the infidelity cost function. 

\subsection{Algorithm robustness}

\begin{figure}[h!]
    \vspace{0.5 cm}
    \includegraphics[width=8.0 cm]{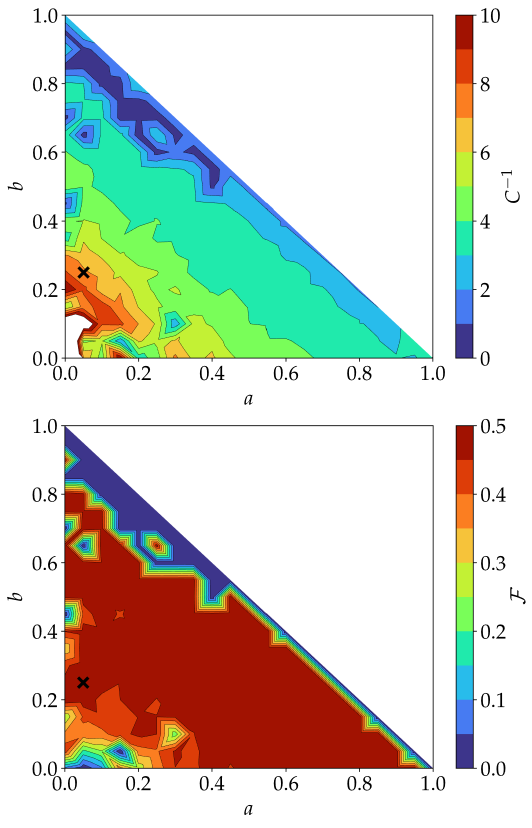}
	\caption{
        Contour plot of the cost function inverse (top) and the fidelity between the scar state and the VQE-S state (bottom) as a function of the cost function parameters $a$, $b$ and $c$.
        The black cross indicates the point chosen in the main text ($a=0.05$, $b=0.25$, $c=0.70$).
        Here we consider $H_1$ and the hardware efficient ansatz with unit depth and 500 iterations.
    }
	\label{fig:mod1ParSweep}
\end{figure}

In the main text, we have shown results for a specific choice of the parameters $a$, $b$, and $c$ in the multi-objective (Pareto) optimization. 
It is natural to wonder how the algorithm behaves as these parameters vary. 
To answer this question, we have done simulations of the model $H_1$ for different values of $a$, $b$, and $c$, under the constraint $a+b+c=1$, targeting $E=0$. 
In this case, we used the HE ansatz with a depth equal to 1; $a$, $b$, and $c$ were varied in steps of $0.05$, and 500 iterations of the cost function optimization were performed for each of their combinations.
In Fig.~\ref{fig:mod1ParSweep} (top panel) we show the contour plot of the cost function inverse as a function of the parameters $a$ and $b$ (from which $c$ is uniquely determined given the constraint $a+b+c=1$). 
We can see that the cost function inverse has a linear gradient in the direction of the bisector $a=b$.
It also has a divergence for $a$ and $b$ going to zero; in this case, the cost function is no longer meaningful for our purposes, and it is easier to reach the convergence.
This graph also shows the importance of the symmetry term of the cost function, without which (\textit{i.e.}, for $a+b=1$) convergence would be impossible.
Furthermore, as we can see from the fidelity (bottom panel of the figure), the convergence is quite good in a broad range of parameter values, so that there is no need for them to be fine-tuned.

\end{document}